\documentclass{article}
\usepackage{ijcai23}

\usepackage{times}
\usepackage{soul}
\usepackage{url}
\usepackage[hidelinks]{hyperref}
\usepackage[utf8]{inputenc}
\usepackage[small]{caption}
\usepackage{graphicx}
\usepackage{amsmath}
\usepackage{amsthm}
\usepackage{booktabs}
\usepackage{algorithm}
\usepackage{algorithmic}
\usepackage[switch]{lineno}
\usepackage{xcolor}
\usepackage{graphicx}
\usepackage{subcaption}
\usepackage{lipsum}  
\usepackage{multirow}
\usepackage{array}
\usepackage{booktabs}
\usepackage{pgfplotstable}
\usepackage{longtable}
\usepackage{subcaption}
\usepackage{todonotes}

\title{Fine-tuning and Prompt Engineering with Cognitive Knowledge Graphs\\ for Scholarly Knowledge Organization}

\author{
Gollam Rabby$^1$
\and
Sören Auer$^1$$^2$\and
Jennifer D'Souza$^{2}$\And
Allard Oelen$^2$
\affiliations
$^1$L3S Research Center, Leibniz University Hannover, Hanover, Germany\\
$^2$Leibniz Information Centre for Science and Technology, Hannover, Germany
\emails
\ gollam.rabby@l3s.de,
\{auer, jennifer.dsouza, allard.oelen\}@tib.eu
}

\begin{document}

\maketitle

\begin{abstract}
    The increasing amount of published scholarly articles, exceeding 2.5 million yearly, raises the challenge for researchers in following scientific progress. Integrating the contributions from scholarly articles into a novel type of cognitive knowledge graph (CKG) will be a crucial element for accessing and organizing scholarly knowledge, surpassing the insights provided by titles and abstracts. This research focuses on effectively conveying structured scholarly knowledge by utilizing large language models (LLMs) to categorize scholarly articles and describe their contributions in a structured and comparable manner. While previous studies explored language models within specific research domains, the extensive domain-independent knowledge captured by LLMs offers a substantial opportunity for generating structured contribution descriptions as CKGs. Additionally, LLMs offer customizable pathways through prompt engineering or fine-tuning, thus facilitating to leveraging of smaller LLMs known for their efficiency, cost-effectiveness, and environmental considerations. Our methodology involves harnessing LLM knowledge, and complementing it with domain expert-verified scholarly data sourced from a CKG. This strategic fusion significantly enhances LLM performance, especially in tasks like scholarly article categorization and predicate recommendation. Our method involves fine-tuning LLMs with CKG knowledge and additionally injecting knowledge from a CKG with a novel prompting technique significantly increasing the accuracy of scholarly knowledge extraction. We integrated our approach in the Open Research Knowledge Graph (ORKG), thus enabling precise access to organized scholarly knowledge, crucially benefiting domain-independent scholarly knowledge exchange and dissemination among policymakers, industrial practitioners, and the general public.
\end{abstract}

\section{Introduction}

In recent years, we saw a steeply rising popularity of Large Language Models (LLMs) for a variety of applications in natural language understanding and generation, content creation, programming assistance, translation, etc.
However, for many applications, also a number of challenges with respect to the application of LLMs became apparent. Besides potential bias and intransparency these include in particular: (a) the limited \textit{context} information that can be exploited by the LLM, (b) \textit{confabulation}, where the LLM generates information or narratives that are plausible-sounding but factually incorrect or misleading, and (c) difficulty in handling specific highly \textit{specialized or niche domains} where training data is limited or the language used is very specific or technical.

These issues are particularly pressing for applications in scholarly communication, i.e., the representation, organization, exchange, and usage of scholarly knowledge. 
Traditionally, scholarly knowledge is represented primarily in scientific articles of which several hundred million are already available and approx. 2.5 million are furthermore added every year.
While LLMs can and are being trained with scientific articles~\cite{jungherr2023using,leegemini} scholarly communication is inherently complex, involving intricate processes and specialized knowledge. 
LLMs struggle to capture the nuances and depth required in academic discourse~\cite{asher2023limits}. 


In this work, we address the context and specialized knowledge issues of LLMs with a Neuro-Symbolic approach intertwining two complementary methods -- (1) fine-tuning LLMs with background knowledge obtained from knowledge graphs and (2) injecting query-specific context knowledge into the prompt.
For both strategies, we leverage a novel type of contextualized knowledge graphs -- cognitive knowledge graphs, which organize knowledge not only as entities and relationships but also in small reusable cognitive units.
A cognitive knowledge graph is a knowledge graph equipped with an overlay structure, which determines reusable patterns (so-called graphlets), that represent common cognitive information structures, such as research contributions comprising entities such as the tacked research problem, the approach, the evaluation, etc.

We leverage cognitive knowledge graphs for (a) fine-tuning existing base models with scholarly knowledge obtained from a cognitive knowledge graph and (b) injecting contextual knowledge from the CKG into the prompt in order to exploit additional context during inferencing.

We evaluate our approach with a comprehensive set of experiments with four different LLMs, viz. Llama 2 (7B and 13B model variants), Mistral (7B), and finally Gemini Pro. 
Our subsequent model evaluations follow a two-fold methodology: 1) automatic evaluations using GPT as an evaluator, and 2) manual evaluations using a human expert evaluator. 
We observe that tasks relying on sparse background knowledge such as from the scholarly domain significantly benefit from injecting contextual knowledge from a CKG into the prompt, especially the research field prediction task. 

In summary, the contributions of this work comprise:
\begin{enumerate}
    \item the definition of the notion of cognitive knowledge graphs, which are capable of capturing contextual knowledge to bridge between neural and symbolic processing as well as human curation,
    \item the conceptualization and implementation of a method for injecting contextual knowledge into prompting as well as fine-tuning,
    \item a comprehensive empirical evaluation of the method with four different LLM with human and LLM assessment.
\end{enumerate}


\begin{figure*}[tb]
    \centering
    \includegraphics[trim=2em 2.8em 2.8em 2em, clip,width= 0.97 \textwidth]{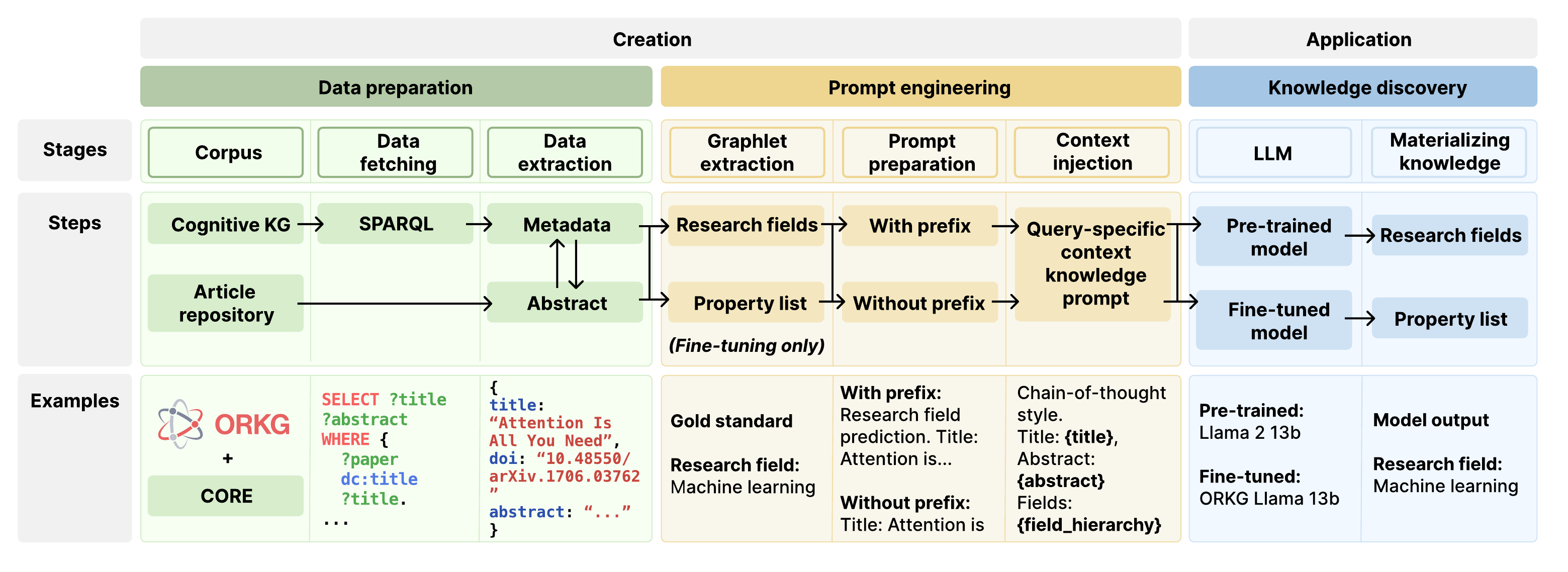}
    \caption{Overview on the method for knowledge augmentation and discovery comprising context knowledge prompt injection and fine-tuning leveraging a Cognitive Knowledge Graph.}
    \label{fig:method}
\end{figure*}



\section{Related Work}

The related work is roughly categorized along the dimensions of Knowledge Graphs, Prompt Engineering for LLMs, and fine-tuning LLMs for specific tasks.

\subsection{Knowledge Graphs}

    

Knowledge graphs primarily focus on representing explicit facts and relationships between entities, often lacking the depth to capture complex, abstract concepts, or contextual aspects~\cite{choudhary2023complex}. They have limitations in representing evolving or multifaceted domains due to challenges in dynamic information updates and integrating multidisciplinary knowledge. These limitations reduce the effectiveness especially for capturing the evolving nature of real-world systems and handling complex information domains, such as scholarly communication. Knowledge graphs are mostly universal and stationary distributed, which does not consider the dynamic nature of many real-world domains~\cite{wang2023dynamic}. Additionally, knowledge graphs face difficulties in updating information in real-time, which is required as domains evolve over time~\cite{hou2023answers}. Furthermore, integrating multidisciplinary knowledge into knowledge graphs is challenging, limiting their ability to represent diverse domains effectively~\cite{peng2023knowledge}. With CKG, we address knowledge graph limitations with dynamic updates and multidisciplinary integration and also enhance their effectiveness in representing evolving or multifaceted domains.

\subsection{Traditional Prompt Engineering: Effective But Limited}

Traditional prompt engineering methods, such as Zero-Shot~\cite{wei2021finetuned}, Few-Shot~\cite{brown2020language}, Chain-of-Thought (CoT)~\cite{wei2022chain}, Tree-of-Thoughts (ToT)~\cite{yao2023tree} have shown improvements in leveraging LLMs for knowledge extraction, fine-tuning, and text generation. However, these approaches do not address limitations arising during the application to low-resource and new domains. 
While deviating from traditional pipelines and offering unique approaches in specific settings like scholarly knowledge graphs~\cite{d2023evaluating}, these methods suffer from two key shortcomings:
\begin{itemize}
    \item \textit{Domain knowledge capture gap:} Out-of-the-box LLMs generally underperform in new domains, indicating the inability to capture relevant domain knowledge~\cite{hu2023prompt}.
    \item \textit{Prompt design constraints:} Identifying appropriate prompt templates requires domain expertise and can be time-consuming, creating a bottleneck in the application~\cite{hu2023prompt}.
\end{itemize}
To address these issues within the scholarly domain, we propose a novel knowledge-driven prompt engineering approach. 
This approach injects query-specific context knowledge from a domain-specific CKG directly into the prompts. 
This not only bridges the knowledge capture gap but also alleviates the need for readily available domain experts and reduces the time required for crafting effective prompts.



    

\subsection{Fine-Tuning: From Prefix Tweaks to Parameter-Efficient Powerhouses}

While prompt-based training and fine-tuning have demonstrably enhanced transformer-based LLM performance (up to 40\% increase under relaxed settings~\cite{d2023evaluating}), different approaches offer varying advantages and limitations. 
Prefix-based tuning, adding trainable tokens to sequences, excels for short and moderate lengths but falters on longer passages~\cite{li2023prefix}. 
Prefix-propagation builds on this concept, achieving superior performance on long documents with 50\% fewer parameters~\cite{zhao2023cpet}. 
However, parameter-efficient fine-tuning methods like adapter-based PEFT bypass training the entire LLM, focusing on tweaking external parameters. 
This surprisingly yields comparable or even better results in downstream tasks~\cite{hu2023llm}, offering a cost-effective and accessible alternative for practical applications.



\section{Method}

\begin{figure}[t]
    \centering
    \includegraphics[width=\linewidth]{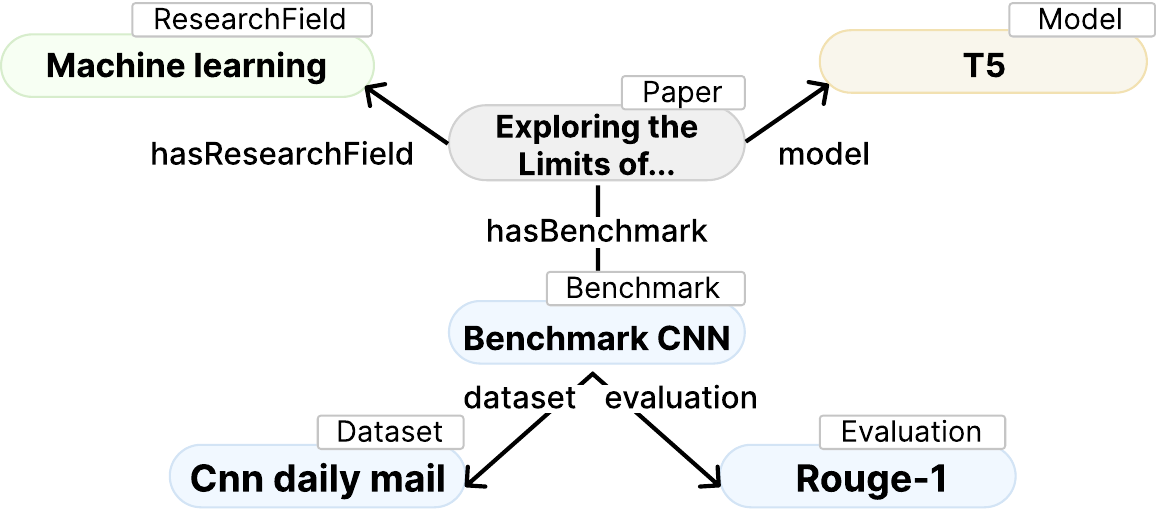}
    \caption{Example of a graphlet retrieved from the ORKG, displaying a scholarly article, the metadata, and the respective properties and entities (simplified version).}
    \label{fig:graphlet}
\end{figure}

Our method comprises three core elements: the novel notion of cognitive knowledge graphs (CKGs), injecting query-specific context knowledge from CKGs into the prompts, and finally fine-tuning LLMs with CKG knowledge.

To illustrate our method, let us consider a scholarly article titled ``Exploring the Limits of Transfer Learning with a Unified Text-to-Text Transformer''~\cite{raffel2020exploring}, for which we aim to perform the following two tasks: (1) predicting the research field and (2) recommending a list of predicates. 
In this scenario, the CKG Open Research Knowledge Graph (ORKG)~\cite{jaradeh2019open} holds metadata and a CKG graphlet for this scholarly article\footnote{\url{https://orkg.org/paper/R161808}}. A visual representation of this specific graphlet is depicted in \autoref{fig:graphlet}. 
The graphlet is essentially a set of typed entities connected through properties such as \texttt{hasTitle}, \texttt{followsMethodology}, \texttt{usesDataset}, \texttt{hasContribution}, \texttt{belongsToResearchField} etc. 
Using ORKG as a CKG, we retrieve this graphlet via SPARQL. 

Simultaneously, we improve the prompts by incorporating real-world context, leveraging abstracts from the CORE dataset. 
This enriches the LLM understanding of scholarly articles and their properties. 
For prompt engineering in this scholarly article, we use the existing CoT prompt framework, injecting it with the extracted query-specific context knowledge for the research field prediction task. 
We include the title, abstract, and research field hierarchy to guide the LLM. 
Additionally, we examine the impact of task-aware prefixes, such as ``Research\_field\_prediction'', to monitor the LLM for specific tasks. 
Integrating CKG graphlets into prompts enhances the LLM's performance for certain cases without needing additional fine-tuning. 
For instance, using the pre-trained Llama 2 13b, this scholarly article can accurately be assigned to the research field of ``machine learning''.
Furthermore, our methodology involves fine-tuning to optimize the LLM's performance, especially in scenarios where prompts lack query- and domain-specific knowledge for a specific task. 
In the case of this scholarly article, without fine-tuning, it is not possible to determine the most suitable domain-specific properties for describing the research contribution, such as \texttt{usesTrainingCorpus}, \texttt{usesTokenization}, \texttt{hasNumberofParameters} etc. 
Through CKG-based fine-tuning, like ORKG\_Llama\_2\_13b, the LLM can extract both domain-specific and domain-independent properties, including \texttt{followsMethodology}, \texttt{usesDataset}, \texttt{hasContribution}, \texttt{usesTrainingCorpus}, \texttt{usesTokenization}, \texttt{hasNumberofParameters}, \texttt{hasModelFamily}, \texttt{hasLicense} etc. 
This integrated approach demonstrates the effectiveness of our methodology in improving the LLM's performance across various scholarly tasks.

\subsection{Cognitive Knowledge Graphs}

Traditional knowledge graphs primarily focus on representing explicit facts and relationships between entities, often lacking the depth to capture complex, abstract concepts or contextual nuances. 
Additionally, they struggle with dynamic information updates and integrating multidisciplinary knowledge, limiting their effectiveness in representing evolving or multifaceted domains, such as scholarly communication.

Our concept of a Cognitive Knowledge Graph now builds upon traditional knowledge graphs by integrating an overlay structure that identifies and utilizes reusable patterns, referred to as graphlets. These patterns represent common cognitive information structures. The motivation for the development and use of CKGs in addition to traditional knowledge graphs is multifaceted.
Among others, we aim for enhanced representation of complex concepts like research methodologies, arguments, or narratives. CKGs, with their overlay structures, can better represent these multifaceted and complex concepts. In particular, we envision CKGs to capture and represent the context in which information exists thus facilitating the moderation between machine and human intelligence. This is particularly important in domains like research, where understanding the context (e.g., the problem addressed, methodologies used, and the nature of the conclusions) is crucial for accurate interpretation.

The base constituents of cognitive knowledge graphs are more complex fabrics of entity descriptions arranged according to certain patterns -- \textit{graphlets}. 
In network analysis and graph theory, the notions graphlet~\cite{10.1093/bioinformatics/bth436,10.1093/bioinformatics/btl301} and motif~\cite{Milo824} were introduced to provide a structuring element between whole graphs and individual nodes and edges. 
Hence, in order to be able to effectively represent and manage more complex knowledge artefacts, we translate and apply the notion of graphlets to knowledge graphs.

Formally, a \textit{CKG graphlet} is a tuple of sets of types (classes) and roles (properties) $(C,P)$, where: 
\begin{enumerate}
    \item for each role $p \in P$, the domain (either explicitly defined or implicitly inferred from a concrete CKG) includes at least one of the types $c \in C$: $domain(p)\subset C$ and 
    \item all types $c \in C$ are connected via a property chain in $P$:  $\forall c_1,c_2 \in C, \; \exists p_1,...,pj,...,p_n \in P: domain(p_1)=c_1 \wedge range(p_n)=c_2 \wedge \bigwedge\limits_{i=1}^{n-1} range(p_i)=domain(p_{i+1})  $
\end{enumerate}
Alternatively, we can also view CKG graphlets as (a) a special type of connected graph patterns (according to the SPARQL algebra), where variables occur in the positions of concrete instances and literals, or (b) as specific sets of SHACL shapes.

\subsection{Injecting Query-specific Context Knowledge into the Prompts}



Our method is based on injecting query-specific context knowledge alongside task-specific context into LLMs prompts to empower them for tasks with sparse knowledge in the foundational models.
We exploit the unique capabilities of CKGs, which represent knowledge not just as isolated facts but as interconnected patterns called graphlets. 
These graphlets can capture more complex concepts and nuanced relationships, providing a richer scaffolding for the LLM's reasoning process. 
Our rationale is to inject query-specific contexts and verified task-driven knowledge with a structured hierarchy from a CKG, such that when both are incorporated into the prompts, the LLMs can achieve increased performance on tasks with sparse prior knowledge or limited training without requiring further fine-tuning. 

For this experiment, we utilized a multi-stage pipeline (cf. \autoref{fig:method}) to inject query-specific context knowledge into the prompts:

\begin{enumerate}
    \item \textit{CKG Knowledge Extraction:} We tap into the vast knowledge reservoir of a CKG, extracting relevant graphlets that align with the specific scholarly task. This goes beyond simple fact retrieval, capturing the intricate cognitive structures and relationships within the domain.
    
    \item \textit{Contextual Grounding:} We leverage abstracts from the CORE dataset~\cite{Knoth2023-zi} as factual summaries, providing rich contextual grounding for the subsequent prompt construction.
    
    \item \textit{Knowledge-Infused CoT Prompts:} The existing CoT prompt framework, a proven way to guide the LLM's reasoning, is enriched with the extracted CKG graphlets. This new step aims to craft prompts that not only guide the LLM's thought process but also equip it with the precise domain-specific knowledge needed for accurate and insightful responses.
    
    \item \textit{Targeted Knowledge Retrieval:} Precise SPARQL queries act as filters, sifting through the vast CKG and extracting only the most relevant information pertinent to each specific prompt. This ensures that the injected knowledge directly aligns with the task.
    
    \item \textit{Task-Aware Prefixes:} To explore the impact of task identification, we utilized optional prefixes attached to the knowledge-driven prompts. These prefixes explicitly signal the scholarly domain and task nature, thus potentially further enhancing the LLM's ability to focus its reasoning and deliver even more refined responses.
\end{enumerate}

By systematically constructing this knowledge-driven pipeline, we aim to improve the performance of LLMs for complex tasks with sparse domain knowledge. 
Injecting query-specific context enriched with structured knowledge from CKGs empowers LLMs to achieve increased performance without the need for additional fine-tuning. 
This approach harnesses the power of LLMs while equipping them with the precise domain-specific knowledge they need to excel in challenging domains with sparse domain knowledge such as scholarly communication.




    \begin{table}[tb]
        \centering
        \begin{tabular}{lccc}
            \toprule
            & \textbf{Train Data} & \textbf{Test Data} \\
            \midrule
            Research Field & 1,894 & 100 \\
            List of Predicates & 1,740 & 100 \\
            \bottomrule
        \end{tabular}
        \caption{Evaluation dataset comprising research field annotations and predicates for research contribution descriptions.}
        \label{tab:dataset}
    \end{table}

    \begin{table}
    \centering
    \begin{tabular}{c|c}
        \toprule
        \textbf{Prompt Style} & \textbf{MAS (RF)} \\
        \midrule
        Zero-Shot Prompting & 67\%  \\
        Few-Shot Prompting & 68\%  \\
        Chain-of-Thought Prompting & 69\% \\
        Zero-shot COT Prompting & 73\% \\
        IQCK into COT Prompting & \textbf{76\%}  \\
        \bottomrule
    \end{tabular}
    \caption{Comparison of our Injecting Query-specific Context Knowledge (IQCK) method with four baseline methods (for a sample set of the CKG graphlets) for the research field prediction task (RF) with Mistral 7b. CoT shares the same input as Few-Shot, with the only difference being the inclusion of additional examples; MAS = Mean Average Score.}
    \label{tab:prompt_Comparison}
    \end{table}

\subsection{LLM Fine-tuning}

In order to increase the LLM performance, we intertwine prompt knowledge injection and fine-tuning leveraging CKGs, specifically in scenarios where the LLM lacks domain-specific knowledge for a task. 
Enriching prompts with query-specific context is powerful, but for domain-specific tasks, injecting additional domain-specific knowledge through fine-tuning might further increase performance. 
This process not only supplements the LLM knowledge but integrates its interaction, expanding its capabilities, especially for domain-specific tasks. 
When faced with domain knowledge gaps, targeted fine-tuning with external domain-specific knowledge proves effective in unlocking the full potential of the LLM. 
Domain-specific knowledge from a knowledge base like CKG, verified by domain experts, equips the LLM for navigating domain-driven challenges. The LoRA~\cite{hu2021lora} attention mechanism efficiently handles intricate details in domain-specific knowledge, ensuring focused involvement. To optimize the resource constraints, quantization~\cite{xu2023qa} is utilized, and the TRL trainer~\cite{vonwerra2022trl} orchestrates the fine-tuning process by setting optimal parameters to overcome learning challenges.

\section{Experimental Setup}



\begin{table*}
  \centering
  \caption{AI (gpt-4-1106-preview) Perspective on LLM (Pre-trained and Fine-tuned) Evaluation; MAS = Mean Average Score; LP = List of predicates recommendation, RF = Research field prediction.}
  \label{tab:example-ai}
  
  \begin{subtable}{0.5\linewidth}
    \centering
    \caption{Table 2.1: AI Perspective on LLM (Pre-trained) Evaluation.}
    \label{tab:pretrain}
    \begin{tabular}{ccccl}
      \toprule
      \textbf{Task} & \textbf{Prefix} & \textbf{LLM} & \textbf{MAS} \\
      \midrule
      \multirow{4}{*}{LP} & \multirow{4}{*}{Without Prefix} & Gemini pro & \textbf{64\%} \\
      & & Llama 2 13b & \textbf{64\%} \\
      & & Mistral 7b & \textbf{64\%} \\
      & & Llama 2 7b & 59\% \\
      \midrule
      \multirow{4}{*}{LP} & \multirow{4}{*}{With Prefix} & Llama 2 13b & \textbf{65\%} \\
      & & Mistral 7b & 64\% \\
      & & Llama 2 7b & 63\% \\
      & & Gemini pro & 62\% \\
      \midrule
      \multirow{4}{*}{RF} & \multirow{4}{*}{Without Prefix} & Mistral 7b & \textbf{79\%} \\
      & & Gemini pro & 78\% \\
      & & Llama 2 13b & 62\% \\
      & & Llama 2 7b & 50\% \\
      \midrule
      \multirow{4}{*}{RF} & \multirow{4}{*}{With Prefix} & Gemini pro & \textbf{80\%} \\
      & & Mistral 7b & 77\% \\
      & & Llama 2 13b & 61\% \\
      & & Llama 2 7b & 43\% \\
      \bottomrule
    \end{tabular}
  \end{subtable}%
  \begin{subtable}{0.5\linewidth}
    \centering
    \caption{Table 2.2: AI Perspective on LLM (Fine-tuned) Evaluation.}
     \label{tab:finetune}
    \begin{tabular}{ccccl}
      \toprule
      \textbf{Task} & \textbf{Prefix} & \textbf{LLM} & \textbf{MAS} \\
      \midrule
      \multirow{3}{*}{LP} & \multirow{3}{*}{Without Prefix} & ORKG\_Llama\_13b & \textbf{67\%} \\
      & & ORKG\_Llama\_2\_7b & 42\% \\
      & & ORKG\_Mistral\_7B & 42\% \\
      \midrule
      \multirow{3}{*}{LP} & \multirow{3}{*}{With Prefix} & ORKG\_Llama\_13b & \textbf{57\%} \\
      & & ORKG\_Mistral\_7B & 52\% \\
      & & ORKG\_Llama\_2\_7b & 37\% \\
      \midrule
      \multirow{3}{*}{RF} & \multirow{3}{*}{Without Prefix} & ORKG\_Llama\_2\_7b & \textbf{42\%} \\
      & & ORKG\_Mistral\_7B & 38\% \\
      & & ORKG\_Llama\_13b & 13\% \\
      \midrule
      \multirow{3}{*}{RF} & \multirow{3}{*}{With Prefix} & ORKG\_Llama\_13b & \textbf{49\%} \\
      & & ORKG\_Mistral\_7B & 46\% \\
      & & ORKG\_Llama\_2\_7b & 19\% \\
      \bottomrule
    \end{tabular}
  \end{subtable}
\end{table*}

\begin{table*}
  \centering
  \caption{Human Perspective on LLM (Pre-trained and Fine-tune) Evaluation; MAS = Mean Average Score; LP = List of predicates recommendation, RF = Research field prediction.}
  \label{tab:human_eval}
  
  \begin{subtable}{0.5\linewidth}
    \centering
    \caption{Table 3.1: Human Perspective on LLM (Pre-trained) Evaluation.}
    \begin{tabular}{cccccc}
      \toprule
      \textbf{Task} & \textbf{Prefix} & \textbf{LLM} & \textbf{MAS} \\
      \midrule
      \multirow{4}{*}{LP} 
      & \multirow{4}{*}{With Prefix} & Llama\_2\_13b & \textbf{60\%} \\
      & & Mistral\_7b & \textbf{60\%} \\
      & & Llama\_2\_7b & 59\% \\
      & & Gemini\_pro & 57\% \\
      \midrule
      \multirow{4}{*}{LP}
      & \multirow{4}{*}{Without Prefix} & Llama\_2\_13b & \textbf{60\%} \\
      & & Llama\_2\_7b & 59\% \\
      & & Mistral\_7b & 59\% \\
      & & Gemini\_pro & 58\% \\
      \midrule
      \multirow{4}{*}{RF} 
      & \multirow{4}{*}{With Prefix} & Gemini\_pro & \textbf{78\%} \\
      & & Mistral\_7b & \textbf{78\%} \\
      & & Llama\_2\_13b & 70\% \\
      & & Llama\_2\_7b & 49\% \\
      \midrule
      \multirow{4}{*}{RF}
      & \multirow{4}{*}{Without Prefix} & Mistral\_7b & \textbf{82\%} \\
      & & Gemini\_pro & 75\% \\
      & & Llama\_2\_13b & 66\% \\
      & & Llama\_2\_7b & 56\% \\
      \bottomrule
    \end{tabular}
  \end{subtable}%
  \begin{subtable}{0.5\linewidth}
    \centering
    \caption{Table 3.2: Human Perspective on LLM (Fine-trained) Evaluation.}
    \begin{tabular}{cccc}
      \toprule
      \textbf{Task} & \textbf{Prefix} & \textbf{LLM} & \textbf{MAS} \\
      \midrule
      \multirow{3}{*}{LP} 
      & \multirow{3}{*}{With Prefix} & ORKG\_Mistral\_7B & \textbf{53\%} \\
      & & ORKG\_Llama\_13b & 50\% \\
      & & ORKG\_Llama\_2\_7b & 45\% \\
      \midrule
      \multirow{3}{*}{LP} 
      & \multirow{3}{*}{Without Prefix} & ORKG\_Llama\_13b & \textbf{60\%} \\
      & & ORKG\_Llama\_2\_7b & 47\% \\
      & & ORKG\_Mistral\_7B & 46\% \\
      \midrule
      \multirow{3}{*}{RF} 
      & \multirow{3}{*}{With Prefix} & ORKG\_Llama\_13b & \textbf{50\%} \\
      & & ORKG\_Mistral\_7B & 48\% \\
      & & ORKG\_Llama\_2\_7b & 33\% \\
      \midrule
      \multirow{3}{*}{RF} 
      & \multirow{3}{*}{Without Prefix} & ORKG\_Mistral\_7B & \textbf{55\%} \\
      & & ORKG\_Llama\_2\_7b & 41\% \\
      & & ORKG\_Llama\_13b & 18\% \\
      \bottomrule
    \end{tabular}
  \end{subtable}
\end{table*}

\paragraph{Dataset.}
Our dataset originates from a large-scale CKG, the Open Research Knowledge Graph (ORKG), encompassing diverse research fields and a comprehensive list of predicates. 
To enhance query-specific content injection and facilitate fine-tuning, we utilized 1,894 and 1,740 graphlets related to the research fields and a list of predicates (\autoref{tab:dataset}), respectively. 
This dataset creates a knowledge-rich environment for the LLMs, with a comprehensive understanding of various academic fields. 
Additionally, we incorporated 100 graphlets in the test set, each carefully curated to showcase a list of research fields and list of predicates. 
These graphlets encapsulate the essence of their respective fields, emphasizing key concepts with a list of predicates. 
The curated CKG graphlets dataset is employed to unveil the true potential of LLMs when faced with intricate scholarly tasks.
The choice of 100 graphlets for the test set serves multiple purposes. 
It not only enables us to assess the LLM's performance through automated-generated metrics but also facilitates a rigorous human validation process~\cite{shen2023large}~\cite{chew2023llm}. 
By comparing the LLM's generated responses with the judgments of human evaluators, we gain a deeper understanding of its strengths and weaknesses. 
This cross-validation approach, integrating AI and human evaluation, ensures a comprehensive and granular assessment of the LLM's capabilities, ultimately paving the way for its further development and refinement.

\paragraph{Evaluation criteria}
The evaluation criteria encompass two primary dimensions: (1) LLM-based evaluation~\cite{liusie2023zero}~\cite{zhang2023classifying} and (2) human-based evaluation~\cite{wu2023style}~\cite{shen2023large}. 
In LLM-based evaluation, the research field prediction is assessed based on clarity, coverage, relevance to the research field predicted by a domain expert, and granularity of the specified LLM-generated research field, each scored within a range of 0-3. 
Similarly, the LLM's list of predicate recommendations is evaluated considering clarity, coverage, relevance for domain expert recommended contexts for a scholarly article, granularity, and the LLM's ability to recognize all contexts, each scored within the same 0-3 scale. 
Human-based evaluation mirrors LLM-based criteria, ensuring a parallel human assessment of clarity, coverage, relevance, and granularity in both research field categorization and a list of predicate recommendations. 
This approach aims to comprehensively evaluate the LLM's performance, combining automated metrics with human judgment to provide valuable insights and an in-depth understanding of its strengths and weaknesses in scholarly-related tasks.

\paragraph{Tasks for Tuning}

This research explores two different approaches for injecting query-specific knowledge into prompts: task-independent and task-driven variants. 
By employing these two distinct prompt variants in our experiment, we assess the strengths and weaknesses associated with task-independent and task-driven approaches, analyzing how LLMs handle different graphlet structures.
\begin{itemize}
    \item \textit{Task-Independent Prompts:} We extract query-specific context from the CKG without imposing any task-related constraints. LLMs equipped with general knowledge can readily address diverse tasks. Additionally, the LLM undergoes fine-tuning with this general CKG knowledge, further enhancing its understanding of underlying concepts and relationships.
    \item \textit{Task-Driven Prompts:} In the second approach, we harness the power of CKGs to formulate task-specific prompts, explicitly incorporating task-oriented prefixes that guide the LLM's reasoning toward the desired outcome. For instance, in a research field prediction task, the prompt might commence with ``Research field prediction'' along with all the pertinent CKG-derived context. This targeted approach aims to improve the LLM's precision in applying its knowledge to the specific task.
\end{itemize}

\section{Evaluation}
\label{sec:results}


We performed a comparative analysis of different prompt engineering approaches (\autoref{tab:prompt_Comparison}) and LLM performance for both pre-trained (\autoref{tab:pretrain}) and fine-tuned (\autoref{tab:finetune}) models across various tasks, which helps to clarify the capabilities and limitations of these LLMs regarding the CKG knowledge injection approach. 

The comparison of our Injection of Query-specific Context Knowledge (IQCK) method with four baseline methods (for a sample set of the CKG graphlets) for the research field prediction task (RF) with Mistral 7b is shown in \autoref{tab:prompt_Comparison}.
For the baseline methods, we tried different prompting styles but omitted the injection of a research field hierarchy to choose predictions from.
We can observe, that IQCK with the injection of the research field hierarchies into the prompt as contextual knowledge significantly improves the accuracy of the research field prediction to 76\% from the 67-73\% achieved with the baseline methods.

The evaluation of pre-trained LLMs (\autoref{tab:pretrain}), including Gemini\_pro (without and with Prefix) and Mistral\_7b (without and with Prefix) shows impressive mean scores, signifying their capabilities in capturing and leveraging query-specific context knowledge from within the prompts. 
Specifically, Gemini\_pro (with Prefix) achieved a remarkable mean score of 80\% (GPT4 evaluation), positioning it as a top performer in this task. 
Mistral\_7b (without Prefix) follows closely with a substantial mean score of 79\% (GPT4 evaluation) and 60\% (human evaluation), demonstrating robust competency. 
The models Llama\_2\_13b (without and with Prefix) and Llama\_2\_7b (without and with Prefix) achieve lower mean average scores than others by the GPT4 evaluation and are similarly scored by human evaluation. 
This pattern indicates that the incorporation of injecting query-specific context knowledge during the prompt engineering phase provides these LLMs with a solid foundation for understanding and recommending research fields.

The evaluation of fine-tuned LLMs (\autoref{tab:finetune}) in the research field prediction reveals different dynamics. 
For instance, ORKG\_Llama\_2\_13b (With Prefix) achieves a mean score of 49\% (GPT4 evaluation), showing competitive performance with regard to other fine-tuned LLMs but falling short from the previous approach. 
Similarly, ORKG\_Llama\_13b (Without Prefix) and ORKG\_Mistral\_7B (With Prefix) achieve mean scores of 42\% (GPT4 evaluation) and 46\% (GPT4 evaluation), respectively, suggesting a noteworthy but comparatively lower proficiency. 
This variance scores between pre-trained and fine-tuned LLMs underscores the substantial impact of injecting query-specific context knowledge in the pre-train LLMs for research field prediction tasks. 

For the list of predicate recommendations, Llama\_2\_13b (with Prefix) achieved a mean score of 65\% (GPT4 evaluation) and 60\% (human evaluation), emphasizing its proficiency in suggesting predicate labels. 
Llama\_2\_7b (With Prefix) follows closely with a mean score of 63\% (GPT4 evaluation), while Llama\_2\_13b (With Prefix) and Mistral\_7b (With Prefix) achieve mean scores of 63\% (GPT4 evaluation) and 64\% (GPT4 evaluation) which is also quite similar to a human evaluation, respectively.

Regarding the fine-tuned LLMs in the list of predicates recommendation unveils a more diverse scenario. ORKG\_Llama\_2\_13b (Without Prefix) achieves 67\% of the mean average score, suggesting that fine-tuning can yield competitive results in this task. 
However, ORKG\_Llama\_13b (With Prefix) and ORKG\_Mistral\_7B (With Prefix) experience lower mean scores, with 57\% and 52\%, respectively. 
This variability in performance indicates that the effectiveness of fine-tuned models in predicate label recommendation is more contingent on specific training conditions, and adding a prefix seems to introduce additional complexity.

This comprehensive analysis establishes a solid foundation for LLMs, particularly in the scholarly domain for tasks like research field prediction and list of predicate recommendations. 
The injection of query-specific context knowledge into the prompts from CKG improves the LLM performance and also provides a robust understanding of complex relationship prompts and LLMs.  
Fine-tuned LLMs showcase competitive performance, especially for list of predicate recommendations, highlighting the relation between the LLMs and fine-tuned requirements.

\section{Limitations and Future Work}



In order to evaluate our approach, we utilized a domain expert to verify data using the previously described approach. 
Although the domain expert verified data contains human curated data, it can be misleading to determine to performance of our models. 
For the research field prediction task, the domain expert verified data only includes a single field per article. 
In the case of interdisciplinary fields, still, only a single field is contained. 
Consequently, the LLM recommendation can be correct based on the contents of the title and abstract but is evaluated as incorrect because of the domain expert-verified data. 
For the list of predicate recommendations, the domain expert verified data only includes a subset of relevant predicates, since knowledge descriptions in the CKG are generally incomplete. 
Therefore, an LLM-recommended list of properties might be relevant based on the title and abstract, but evaluated as incorrect when comparing them to the domain expert-verified data. 

Furthermore, there are several limitations regarding the human evaluation of the data. 
The evaluators were instructed on how to score the outcomes, but only high-level guidelines were provided. 
Therefore, scores might vary between different articles. 
Despite this shortcoming, the human evaluation still provides valuable insights, especially in comparison with the machine evaluation. 
Another aspect of the human evaluation is the heterogeneous set of research domains. 
Since our approach does not focus on specific domains, the set of selected articles comes from a variety of domains. 
The human evaluators were not experts in all domains and therefore had to judge the results to the best of their ability. 
However, we believe that a high-level assessment of the resulting LLM responses is also possible without domain knowledge, albeit in lesser granularity.  

In addition to the two tasks performed in our approach, we plan to extend this to various other tasks in future work. 
In addition to the prediction of predicates, we explored the prediction of objects to form complete contribution description facts. 
However, further work is required to perform this task at the desired accuracy. In the future, we also plan to continue working on this specific task by leveraging other approaches, such as fine-tuning the LLMs for each specific task and increasing the data from the CKG which needs to be collected from the domain experts.

\section{Conclusion}

This work is part of a larger research agenda aiming at creating a comprehensive neuro-symbolic system for describing research contributions in a cognitive knowledge graph ultimately giving rise to novel AI-based research assistance systems, which enable researchers to obtain comprehensive answers to research questions based on the recent corpus of scientific knowledge.
Due to the limitations of LLMs in extracting information in domains with sparse training data, we developed two methods for (1) injecting contextual information into prompts from a preexisting CKG and (2) fine-tuning LLMs with such knowledge for specific tasks.
Our evaluation showed, that in particular the first method is very well suited to improve the accuracy of LLMs for scholarly communication tasks.
More research is required to further improve the fine-tuning methods and expand the evaluation of the approach to further knowledge extraction and augmentation tasks, such as object prediction.




\bibliographystyle{named}
\bibliography{ijcai23}

\end{document}